\input phyzzx
\sequentialequations
\overfullrule=0pt
\tolerance=5000
\nopubblock
\twelvepoint

\line{\hfill  IASSNS 97/22}
\line{\hfill hep-th/9703131}
\line{\hfill March 1997}

\titlepage
\title{Cross-Confinement in Multi-Chern-Simons Theories}
\author{Lorenzo Cornalba\foot{cornalba@princeton.edu}}
\centerline{{\it Department of Physics}}
\centerline{{\it Joseph Henry Laboratories}}
\centerline{{\it Princeton University}}
\centerline{{\it Princeton, NJ 08544, USA}}
\vskip.4cm
\author{Frank Wilczek\foot{Research supported in part by DOE grant
DE-FG02-90ER40542.~~~wilczek@sns.ias.edu}}
\vskip.2cm
\centerline{{\it School of Natural Sciences}}
\centerline{{\it Institute for Advanced Study}}
\centerline{{\it Olden Lane}}
\centerline{{\it Princeton, N.J. 08540}}
 
\endpage
 
\abstract{We identify a class of 2+1 dimensional models, involving
multiple Chern-Simons gauge fields, in which a form of classical
confinement occurs.  This confinement is not cumulative, but
allows finite mass combinations of individually confined objects, as
in baryons.  
The occurrence and nature
of the phenomena depends on number theoretic properties
of the couplings and charges.}
 
\endpage

In two spatial dimensions, one has an almost trivial mechanism of
confinement: the Coulomb field of a charged particle decays as $1/r$,
and therefore the field energy 
diverges logarithmically at large $r$.  Compared
to confinement in QCD, this mechanism differs notably in two ways: in
QCD confinement the energy of an uncompensated color charge
diverges linearly
rather than logarithmically with distance, and in QCD the confining force is
not simply quadratic in the charge -- indeed, 
suitable multiples of confined charges can form finite-energy states,
as in baryons.  In this note we shall discuss a slightly more
elaborate class of 2+1 dimensional field theories that exhibit this
second feature.  Our field theories are of the multi-Chern-Simons
form, and are closely related to those used in the hierarchical
construction of quantum Hall states.  Perhaps surprisingly, our form 
of confinement phenomenon depends on
number-theoretic properties of the couplings and charges.
For the most part our considerations will be
classical, but we remark briefly on the quantum version toward the end.

Let us first describe the basic mechanism, in its simplest
incarnation, verbally.  Consider a $U(1)_A \times U(1)_B$ gauge theory
including a purely off-diagonal Chern-Simons coupling.  The effect of
this coupling is that an electric charge with respect to one gauge
group induces a magnetic flux with respect to the other.  Suppose
that $U(1)_A$ is in a superconducting phase; specifically,
that there is a scalar field, charged with respect to this group, which
condenses.  The existence of this condensate will quantize the
allowed values of $U(1)_A$ flux.  On the other hand, a particle
charged with respect to $U(1)_B$ will, because of the Chern-Simons
coupling, seek to set up a definite value of the $U(1)_A$ flux,
proportional to the charge.  The `desired' flux value generally will not
respect the quantization condition, however, and in that case the
charge will be confined.  At the same time,  a suitable multiple of  
the confined charge will correspond to an allowed flux value, and it will
not be confined.

\REF\mcs{See D. Wesolowski and Y. Hosotani, and C-L. Ho, {\it
Int. J. Mod. Phys.} {\bf A9}, 969, (1994), and references therein.}

For later use, let us now define the general Lagrangian density of interest
$$
{\cal L} ~=~ 
{1\over 8\pi } \epsilon^{\alpha \beta \gamma}  
\mu^{lm}  a^l_\alpha f^m_{\beta\gamma} 
+ D_\alpha \phi^r  (D^\alpha \phi^r )^*  
- V(\phi ) ~, 
\eqn\basicL
$$
where the covariant derivative on the $r^{\rm th}$ scalar field acts as 
$$
D_\alpha \phi^r \equiv \partial_\alpha \phi^r - i q^r_l a^l_\alpha \phi^r
\eqn\covderiv
$$
(suspending the summation convention on $r$),
and $V$ is an effective potential whose details need not concern
us.  $\mu$ is a symmetric matrix [\mcs ].  We have chosen the normalization in
such a way that integral entries in $\mu$ and integral charges arise
naturally, for example if one demands good global behavior of the
theory on
topologically non-trivial surfaces or that it results from breaking of
an overlying non-abelian symmetry.

At the moment we are
concerned with the simple case when $\mu$  is a 2$\times$2 matrix
whose only
non-zero entries are equal to an integer $n$ off the diagonal.  We assume that
just one scalar field, carrying charges $(0, q)$ with respect to the
two gauge groups, condenses.  The field
equations of most interest to us arise from the variation of $\cal L$
with respect to the time-component of $a^1$ and with respect to the
space-components of $a^2$.  Allowing for an external source carrying 
charge of the first type, these equations read:
$$
{n\over 2\pi } b^2 = \rho^1_{\rm ext.} 
\eqn\fluxeqn
$$
and 
$$
{n\over 2\pi } {\vec e}^1 = \hat z \times {\vec j}^2 ~. 
\eqn\quanteqn
$$
Let us analyze the second of these first.

We require that the condensate lives on the vacuum manifold at
infinity, and is single-valued, so that by a gauge transformation it
can be brought into the form $\phi \rightarrow v e^{il\theta}$,
where $l$
is an integer and $v$ is magnitude of the vacuum expectation value of
$\phi$.  
In order that the current
due to the condensate
fall off faster than $1/r$ at infinity we must further require that
$$
l ~=~ q a^2_\theta ~, 
\eqn\lvsa
$$
in order that the azimuthal covariant derivative vanish.  
Using
Stokes' theorem,
\lvsa\ integrates to the flux quantization condition
$$
2 \pi l ~=~ q \Phi^2 ~.
\eqn\fluxquant
$$
Integrating \fluxeqn\ over all space, on the other hand, gives us
$$
Q^1_{\rm ext.}  ~=~ {n\over 2\pi } \Phi^2~.
\eqn\chargeflux
$$
Combining these, we find the condition
$$
Q^1_{\rm ext.} ~=~ {n l\over q}~.
\eqn\unconfinedq
$$
on the external charge.
If $n$ does not divide $q$, this condition will generally fail, even
for integer $Q^1_{\rm ext.}$.  If this condition  fails,
it will not be possible for the fields in \quanteqn\ to fall off
faster than $1/r$ at infinity.  The resulting long-range fields lead 
to energy diverging 
logarithmically
with distance, proportional to the square of the fractional part of 
${q\over n}Q^1_{\rm ext.}$, arising from the gradient energy of the
condensate, the Coulomb field energy of the electric field (if a Maxwell
term is present in an appropriate, expanded form of $\cal L$), or both.  
Thus charges which do not satisfy \unconfinedq\ are confined.  On the
other hand, clearly an appropriate multiple of a confined charge will
not be confined.  The finite-energy states will include baryon-like
objects, but not the corresponding quarks.

Now let us consider the more general 2$\times$2 case, where  
$$
\mu ~=~ \left(\matrix{m_1 &n \cr n& m_2\cr}\right)
\eqn\mumatrix
$$
and the condensate has charge vector $(q^1, q^2)$.  Following steps similar
to those above, we find conditions
$$
\eqalign{
Q^1_{\rm ext.} ~&=~ {1\over 2\pi} ( m_1 \Phi^1 + n \Phi^2 ) + q^1
\lambda \cr
Q^2_{\rm ext.} ~&=~ {1\over 2\pi}  (n \Phi^1 + m_2 \Phi^2) + q^2
\lambda \cr }
\eqn\chargecond
$$
and
$$ 
2\pi l ~=~ q^1 \Phi^1 + q^2 \Phi^2~,
\eqn\flcond
$$
for the long-ranged fields to vanish.  
Here $\lambda$ is a continuous parameter, representing the ability of
the condensate to screen electric charge.  The ratio of screening 
charges, of
course, must follow that of the condensate.  Now in the generic case,
after solving \flcond , there will be two continuous parameters
$\lambda$ and $\Phi^2$ available to satisfy the two equations
\chargecond , and then an arbitrary charge will be screened (not
confined). However if
$$
m_2(q^1)^2 - 2n (q^1)(q^2) + m_1 (q^2)^2 ~=~ 0 ~,
\eqn\quadcond
$$
then these two parameters multiply proportional coefficients in
\chargecond . Hence there will be charges that cannot be screened, and
must be confined.  (Our earlier case corresponds to $m_1 = m_2 = q^1 =
0$.)  A brief calculation reveals that the condition 
$Q\mu^{-1}q = l$ which must be
satisfied by unconfined charges can be written 
in the transparent form 
$$
q^1Q^2_{\rm ext.} - q^2Q^1_{\rm ext.} = \sqrt{-\Delta} l~, 
\eqn\concond
$$
where the determinant $\Delta \equiv m_1m_2 -n^2$.
{}There will be non-trivial real solutions $(q^1, q^2)$ of \quadcond\
if and only if $\Delta \leq 0$,    
There will be integer solutions if and only if $-\Delta$ is a
perfect square.

In the 3$\times$3 case, if there is one condensate, a condition
analogous to \quadcond\ is necessary for our confinement
phenomenon: $q\mu^{-1} q = 0 $, in an evident vector/matrix notation. 
(If $\Delta = 0$, one must use the analytic expression 
$\Delta \mu^{-1}$ in place of $\mu^{-1}$ in this equation.)
It
is also interesting to
consider the case of two condensates, with charge vectors
$q_A, q_B$.  Then the condition for some discrete charges to escape
screening, and trigger confinement, takes the form
$$
(q_A \times q_B) \mu (q_A \times q_B) ~= 0~. 
\eqn\doublecond
$$
In the general case, a necessary condition for the existence of
charges which are not screened, but do get confined, is that the
determinant of the matrix whose $rs$ element is 
$$
M_{rs} ~=~ q_r \mu^{-1} q_s 
\eqn\neccond
$$ 
vanishes.

It seems to us that the models here discussed are among the simplest
and most tractable to exhibit interesting confinement phenomena.

\REF\book{For overviews of these matters, including reprints of
many of the original papers, see F. Wilczek, {\it Fractional
Statistics and Anyon Superconductivity\/} (World Scientific, 1990),
and M. Stone, {\it  The Quantum Hall Effect\/} (World Scientific, 1992).}

As previously mentioned, the Lagrangians discussed here are
closely related to those employed in effective theories of states of
matter
in
the quantum Hall complex [\book ].  
In considering whether something resembling
our specific
confinement phenomenon could arise in that context, several
interesting problems arise.   The primary one is profound, and its
interest goes beyond these immediate issues: what, if anything, does
it mean to have an uncondensed version of the theory whose condensed,
or `superconducting' version is the fractional quantum Hall effect?
{}From a higher point of view the essence 
of the fractional quantum Hall effect is phase
coherence among electron superfermions -- electrons coupled to a
Chern-Simons field, which imparts effective flux to them.  From that
perspective, it seems
quite logical that there could arise, when the gap closed,
a quasi-metallic state where the
coupling to the Chern-Simons field persists but phase coherence is
lost.  It would resemble a Landau Fermi liquid 
(immersed, unfortunately, in a large magnetic field), but with
additional effective couplings to the Chern-Simons gauge field.  
Insofar as these
couplings are quantized, this state will be discretely different from
the ordinary Fermi liquid theory.  It is a very significant challenge, to
identify a usable diagnostic for this state, or rather family of states.

\REF\polyakov{A. Polyakov, {\it Nuclear Physics\/} {\bf B120}, 429 (1977). } 

Since, in the models discussed here, the gauge fields are effectively
massive, it does not appear that Polyakov's mechanism [\polyakov ] for linear
confinement using monopoles can be realized in them.  For weak
coupling, the long-range field energies discussed here, which do not involve
exponentials of the negative inverse coupling,  would in any case have
dominated up to
very large distances.  The difference reflects the essentially
classical nature of the confinement discussed here.  Because our mechanism
traces back to the cumulative behavior of weak, spatially extended
fields, quantum fluctuations will not much affect it, other than
possibly to renormalize the charge.

{\bf Acknowledgments}

We wish to thank M. de Wild Propitious for a
question which stimulated this investigation.

\refout

\end